# Atomic-scale mapping of interfacial phonon modes in epitaxial YBa$_2$Cu$_3$O$_{7-\delta}$ / (La,Sr)(Al,Ta)O$_3$ thin films: The role of surface phonons

Joaquín E. Reyes González[1], Charles Zhang[2], Rainni K. Chen[2], John Y. T. Wei[2], Maureen J. Lagos[1*]

[1] Department of Materials Science and Engineering, McMaster University, Hamilton, Ontario L8S 3N4, Canada

[2] Department of Physics, University of Toronto, Toronto, Ontario, M5S1A7, Canada

**Abstract**

We investigate the behavior of phonons at the epitaxial interface between YBa$_2$Cu$_3$O$_{7-\delta}$ thin film and (La,Sr)(Al,Ta)O$_3$ substrate using vibrational electron energy loss spectroscopy. Interfacial phonon modes with different degrees of scattering localization were identified. We find evidence that surface contributions from the surrounding environment can impose additional scattering modulation into local EELS measurements at the interface. A method to remove those contributions is then used to isolate the phonon information at the interface. This work unveils interfacial phonon modes in a high-Tc cuprate superconductor, that are not accessible with traditional phonon spectroscopy techniques, and provides a method for probing interfacial phonons in complex oxide heterostructures.

# INTRODUCTION

Since the discovery of high critical-temperature ($T_c$) superconducting cuprates, there have been incessant investigations seeking to unveil the physical mechanism underlying the unconventional superconductivity. Despite early evidence against the primacy of phonons in mediating the Cooper pairing in optimally-doped cuprates, subsequent studies revealed that electron-phonon coupling plays a significant role in the underdoped regime [1,2]. Deepening our understanding of this interaction requires a comprehensive characterization of both bulk and surface phonon modes sustained in superconducting cuprates, particularly in thin-film samples where epitaxial strain is known to reduce the oxygen content [3], and where surface modes might also affect the electron-phonon coupling.

The phonon properties of epitaxial thin films are affected by the film's interaction with the substrate, leading to the generation of new phonon modes at and near the film/substrate interface. To assess this interfacial phonon behavior, local probes with nano-to-atomic scale spatial sensitivity are needed. Furthermore, given the small volumes of matter contained within the thin superconductor films, spectroscopy techniques that are very sensitive to both bulk and surface phonons are imperative to access the entire range of possible phonon excitations in these materials.

Atom wide electron beams fabricated in monochromated scanning transmission electron microscopes (STEMs) are suitable probes for studying vibrational bulk and surface excitations through electron energy loss spectroscopy (EELS) with atomic scale spatial resolution and sub-5meV energy resolution [4]. Recent studies have exploited these capabilities in the investigation of phonon modes at interfaces between semiconductor materials [5–8]. The study on high-$T_c$ superconducting thin films grown epitaxially on substrates is in the nascent stages [9,10]. Studies of interface modes in the most widely studied compounds of the cuprate family, such as $YBa_2Cu_3O_{7-\delta}$ (YBCO), remained to be conducted. For instance, it is known that structural changes to Cu-O chains in YBCO can affect the $T_c$. Therefore, investigating the vibrational response of these structural features near the interfaces could unveil new phonon behavior.

The interpretation of vibrational EELS data from interfaces is not trivial, given that the fast electron can scatter from both bulk and surface phonon modes [11]. Moreso, investigating interface phonon modes requires a careful assessment of surface contributions from the surrounding environment due to effects of long-range surface excitations (e.g., phonon polaritons), which can be captured via dipole scattering processes. To the best of our knowledge the contributions from long-wavelength surface phonon polaritons were not addressed in previous EELS studies of interface phonon modes [6–8,12,13]. In-depth understanding of the physical origin of electron scattering from the interface will have profound implications on our grasp of phonon modes sustained in high-$T_c$ cuprate thin films.

This work reports on the detection of lattice vibrational modes at the interface between a YBCO film and a $(La,Sr)(Al,Ta)O_3$ (LSAT) substrate. Our atomic-scale EELS mapping of YBCO in regions adjacent to the LSAT revealed a combination of delocalized and highly localized sub-nanometer scattering modulation for interfacial modes. Clear differences are visualized within the CuO chain and layers containing the $CuO_2$ planes. Our work demonstrates the importance of assessing the effects of the surrounding environment and calls attention to the spectral contribution from optical surface phonons to the collected data in EELS experiments.

**METHODS**

The YBCO thin films were epitaxially grown onto c-axis LSAT substrates using pulsed laser-ablated deposition (PLD) with a Lambda Physik Complex 201 248 nm KrF excimer laser. Pre-reacted YBCO powder pressed and sintered to >87% density was used as PLD targets. During the growth the laser was operated at a repetition rate of 5 Hz and a fluence of 2 J/cm$^2$ while the substrate was held at 800 °C and in 200 mTorr of $O_2$. The films were post-annealed in $O_2$ at 450 °C and 760 Torr to ensure optimal oxygen doping.

Temperature-dependent resistivity measurements were made with lock-in amplification in the 4-probe geometry, using a Lakeshore AC 370 bridge with an excitation current such that the induced power in the film was less than 1 mW. The temperature was measured with a Lakeshore DT-670 Si diode.

X-ray diffraction (XRD) studies were performed to verify the growth direction and crystalline structure of the YBCO film. We used a Rigaku Smartlab x-ray diffractometer in the $\theta$-$2\theta$ scanning mode. The diffractometer is equipped with a Cu K$_\alpha$ source with a 1.5406 Å wavelength.

For the imaging and spectroscopy work, an electron transparent cross-section sample of the YBCO-LSAT system was fabricated using focused ion beam (FIB) microscopy. This resulted in a suspended YBCO rod attached on the side to a LSAT wedge. We ensure that the sample has uniform thickness and is not warped during the FIB milling. The rod length, width and thickness are about 1 µm, 150 nm and 50 nm, respectively. In this case, the width represents the YBCO film thickness while thickness refers to the section the electron beam penetrates the specimen.

The crystalline structure of the YBCO/LSAT interface was imaged using aberration-corrected STEM. The experiments were conducted on a Nion HERMES 100 equipped with a probe aberration corrector operated at 60 kV, which delivers a spatial resolution of ~1 Å. High-angle annular dark field (HAADF) STEM images were acquired using an electron probe with a 35 mrad convergence semi-angle and beam current of about 17 pA. The projector setting was configured to collect high-angle elastically scattered electrons using an annular detector with inner and outer collection semi-angle of 72 and 200 mrad respectively. These STEM images were acquired with a dwell time of 8 µs and 1024x1024 pixel sampling. An image was obtained by averaging 10 frames thereby improving the signal-to-noise ratio (SNR). All raw datasets were first aligned and then averaged using available tools in Nion swift software.

The vibrational properties were studied using high energy resolution EELS. The experiments were conducted using a Nion HERMES 100 equipped with a monochromator that works in tandem with an Iris spectrometer and a Dectris ELA direct electron detector. The microscope was operated at 60 kV and spatially resolved EELS spectra were acquired using an electron probe with 15 mrad convergence half-angle and a beam current of 1 - 2 pA, . Current was improved at expense of energy resolution performance resulting in an energy resolution of 6 - 7 meV. Inelastic scattered electrons were collected using a circular aperture subtending 11 mrad collection half-angle. This collection conditions allows us to probe phonon modes over the entire Brillouin zone. Thus, both long and short wavelength phonon modes are excited under dipole and impact scattering conditions. The sample was probed at room temperature conditions. The dispersion used for the EELS experiments was 0.83 meV/channel. Single point acquisition consisted of collecting 50 frames with 4 s of acquisition time. For background subtraction, reference EELS spectra were acquired while keeping the probe in vacuum at ~ 5 µm away from the sample under the same acquisition conditions.

Spatially resolved vibrational EELS maps were collected using mapping grids of different sizes (see supplementary material for detailed technical information). To obtain atomic-scale vibrational scattering maps, we used a 30 mrad convergence half-angle electron probe and collected inelastically scattered electrons over 22 mrad half-angle using a 1 mm aperture. We used an acquisition time of 350 ms which allows acquisition of data with high signal-to-noise ratio and minimum sample drift. In similar manner, reference EELS spectra were acquired in vacuum by collecting 30 frames with the same dwell time used for the maps.

To remove the pseudo-elastic contributions from the acquired EELS data, the tails of ZLP of the reference spectrum was used as background curve for subtraction of the corresponding EELS spectrum dataset. This process was successfully used to extract inelastic scattering signal which obeys the principle of detailed balancing [14]. To render a better visualization of the background subtracted EELS, we multiplied it by an ad-hoc factor ($\Delta E^2$, where $\Delta E$ is the energy loss). As a result, the scattering signal towards high energies gets amplified as well as the noise. The amplified noise is preserved in the displayed data and no attempts were performed to suppress it. Further data processing includes the use of a 2D Gaussian filter [15] for energy-integrated vibrational EELS maps.

## RESULTS AND DISCUSSION

### A. YBCO film: Superconductivity and structural properties

Figure 1(a) shows the resistivity of the superconducting YBCO film as a function of temperature and reveals the superconducting transition temperature ($T_c$) to onset around 90 K with a width of 3 K. These values agree with those in other YBCO films grown on LSAT in similar conditions [16,17] and indicate that the bulk of the film is amply oxygenated and thus optimally doped. Figure 1(b) displays the θ-2θ XRD pattern confirming the *c*-axis (001) orientation of the YBCO film. All the prominent features in the black curve are identified as (00ℓ) peaks from the YBCO film or the LSAT substrate. The red curve corresponds to the measurement of the empty sample holder, where the peak labeled with (*) is assigned to a reflection from the material of this sample holder.

A schematic of the fabricated YBCO/LSAT sample is shown in Fig. 2(a). Figure 2(b) shows a STEM-HAADF image of the cross-section view of the epitaxial YBCO film on the LSAT substrate. A closer inspection of this YBCO/LSAT interface (dashed lined) indicates that the atomic arrangement is structurally sharp. The YBCO top surface is rougher and the effective thickness of the YBCO film is about 150 nm. The bright narrow feature over the YBCO film is the tungsten (W) protective layer that was deposited during the TEM sample preparation to prevent ion milling damage.

YBCO crystallizes into an orthorhombic structure, along the *c*-axis the crystal is visualized as a stack of single layer units spaced by CuO chains. The layered unit arrangement is composed of a Y plane separating two adjacent $CuO_2$ planes which are enclosed by two BaO planes. The crystalline structure of epitaxially-grown YBCO films on LSAT substrates was previously studied [16]. Our atomic-scale imaging also revealed regions of double CuO chain intergrowths, which have been previously observed in heteroepitaxial YBCO films and superlattices [3,18]. Spectroscopy measurements were performed while avoiding regions containing microstructural crystalline defects. Therefore, the EELS data contains prominent contributions from crystalline YBCO regions.

**B. Phonon behavior across the YBCO/LSAT interface**

To assess the behavior of phonons at the YBCO/LSAT interface, we probed the vibrational response of the region surrounding the interface. Figure 2(c) shows EELS spectra acquired at discrete positions across the YBCO/LSAT interface, going from the center region of the YBCO towards the LSAT substrate.

On the YBCO side (blue spectra), the EELS spectra acquired at the central region (around 80 nm) comprise two broad spectral features with maxima of around 55 and 82 meV that are spaced by a minor contribution of around 75 meV. The scattering appearing above 82 meV is weak and is associated with the excitation of surface phonon polaritons of the neighbouring LSAT structure. These weak contributions disappear above 100 nm from the interface. The spectral range between 20 and 80 meV is dominated mostly by bulk optical phonon excitations of YBCO, which involve lattice motion of Cu-O ions. Between 35-80 meV, the dominant contribution comes from $CuO_2$ plane motion, while CuO chains and BaO contribute primarily in the 20-35 meV range, with smaller contributions appearing between 45-60 meV [19]. There is a reasonable similarity of the EELS results with the phonon density of states, which suggests the approximation of vibrational EELS spectrum with the partial ground-state phonon density of states [20].

The EELS signal acquired towards the interface shows an apparent shift of the 82 meV peak towards higher energies, accompanied with new spectral features up to about 100 meV, and the 55 meV broad resonance tends to sharpen. The overall scattering intensity increases gradually towards the interface. Very close to the YBCO/LSAT interface (magenta spectrum) strong scattering changes appear in the range of about 40 to 105 meV, which are induced by contributions from both interface and surface phonon modes from the YBCO and LSAT structures. This scattering behavior will be further discussed below. On the LSAT side (red spectra), minor scattering variations are noticed within 10 nm from the interface. These findings reveal the richness of the vibrational scattering within a region of 40 nm in size across the YBCO/LSAT interface.

Towards the interior of the LSAT (above 40 nm) the EELS spectra are dominated by two peaks at 58 meV and 92 meV, which are accompanied by a small shoulder-like spectral feature at 38 meV. The scattering within this spectral range is associated with the excitation of bulk phonon modes of LSAT. Sr-BO$_6$ (B=Al, Ta) motions have been measured at 35 and 44 meV, while B-O bending and stretching optical phonons lie in the 48-68 meV and 78-100 meV ranges respectively [21].

## C. Interface phonons: The role of surface phonon modes

As mentioned above, the phonon scattering signal collected at the interface contain both localized and delocalized scattering contributions. Delocalized contributions can arise from the excitation of localized surface phonons (plasmons) sustained in dielectric (metallic) structures of finite size [22], such as a YBCO rod, LSAT wedge or W rod. For instance, localized surface phonon modes sustained in YBCO rods can be excited with different scattering probabilities by a fast electron travelling in aloof mode, as shown in Fig. S4. These long wavelength polaritonic modes are excited via dipole scattering processes ($q \rightarrow 0$) and their mode energies lie within the Reststrahlen bands of the material. Similar response is expected for a LSAT wedge excited by an electron probe in aloof mode. Furthermore, one might consider the generation of coupled phonon-phonon and phonon-plasmon surface modes due to the interaction of the YBCO rod with the LSAT support and the W layer respectively. In this section, we focus on analysing the delocalized contributions from those surface excitations to the signal acquired at the YBCO/LSAT interface (Figure 2).

Figure 3(a) shows a probe position dependent 2D EELS spectra acquired across the YBCO structure, which highlight the richness and complexity of the EELS scattering. We observe that the scattering distribution exhibits an asymmetric profile with respect to the YBCO central region and gradually increases towards the YBCO edges. Increase in scattering is expected towards the YBCO edges due to contributions from surface phonons at the expense of bulk contributions, as dictated by the Begrenzung effect [11,23]. This scattering behavior is expected to be symmetric, however this symmetry profile can be disrupted by scattering contributions from both interface phonons and surface phonons of the neighboring LSAT structure. We observed that the surface contribution from YBCO increases rapidly within 10 nm from the YBCO/W edge, and a similar response is expected on the opposite interface. A close inspection of the YBCO/LSAT interface reveals that in addition to the YBCO surface contribution there are continuous scattering transitions across the interface at particular energies, which suggest the presence of active interfacial phonon modes. Furthermore, contributions from LSAT surface phonon polaritons spread into the regions occupied by the YBCO, as illustrated by the scattering within the 85-100 meV band.

It is important to mention that the scattering enhancement towards the YBCO/W interface does not contain contributions from coupled phonon-plasmon modes. Those hybrid excitations are not generated due to a large detuning ($E_{plasmon} \gg E_{phonon}$) between their energy modes [22]. In a similar manner, surface phonon-phonon coupling between the YBCO rod and LSAT structure are expected to be weak due to detuning between their surface phonon modes. Therefore, those two effects are not considered in this study as their contributions are negligible.

Isolating the interfacial phonon modes requires removing the delocalized surface scattering contributions from both the YBCO and LSAT structures. To address the YBCO contributions, we use an approach that capitalizes on the spatial symmetry of the surface scattering profile across the YBCO rod and the physics of the Begrenzung effect. Very close to the YBCO/W interface, the spectrum is dominated by surface phonon components with minimal contributions from the bulk [11]. Due to symmetry arguments previously stated, similar surface contributions are expected on both sides of the YBCO, thus this contribution can be removed by performing a subtraction between a spectrum acquired near the YBCO/LSAT interface and its counterpart at an equidistant position on the opposite side (Fig. S1(a)). The resulting spectra obtained near the interface and at the middle is shown in Fig. 3(b). The spectra acquired near the interface shows two broad bumps centered at around 55 and 85 meV separated by a small shoulder-like spectral feature around 75 meV. Acquisitions with larger exposure time reveals the same spectral features with lower noise level (bottom spectrum in Figure 3(b)). Interestingly, we found that the resulting spectra for acquisition at farther distances from the interface (e.g. 10 nm) still preserve the same spectral characteristics. At the middle region (e.g. 50 nm – 75 nm) the resulting spectra (top of Figure 3(b)) approaches zero due to similarity of the scattering. It is important to notice that the validity of this method is limited to regions close to the interface.

We then address the LSAT contributions. To assess them accurately is challenging but it is possible to get a good estimate by considering two key aspects grounded on the physics of the inelastic electron scattering process: Measure an EELS spectrum from an isolated LSAT wedge in the aloof mode with the same impact parameters and structure geometry used for the experiments probing across the YBCO/LSAT interface (Fig. 3(a)). Then consider screening effects of the YBCO on the travelling electron to estimate the magnitude of the spectral component of the LSAT contribution to be subtracted. This procedure is described in detail in the supplementary material. The intensity of the surface scattering from the neighboring LSAT structure is more prominent towards the ends of the structure under examination than towards the center. This variation profile was also observed in a polar dielectric wedge attached to a phononic rod [22]. The current analysis was done at a point where the LSAT contributions are most intense so the effect of the delocalized surface contributions can be easily visualized and its significance appreciated. For this case study, the aloof LSAT contributions can still be detected halfway into the YBCO region (around 80 nm), but interface regions can be strategically selected where LSAT contributions are weaker.

Once we determine the LSAT contributions we proceed to its removal. The green curve of Figure 3(c) shows the resultant subtracted spectrum which is dominated by two broad peaks centered at about 40 and 75 meV. A comparison with the original EELS spectrum acquired at the interface (magenta curve in Figure 2b) clearly reveals the contribution from the delocalized surface scattering, highlighting the importance of accounting the surrounding environment on vibrational mode studies. These collective surface contributions accounted for up to about 60-70%% of the original scattering in our case study, imposing a strong modulation on the spectrum. Moreover, a comparison between the spectrum for YBCO/LSAT interface (green curve), YBCO (blue curve) and LSAT (red curve) reveals distinct scattering profiles and points out to the existence of interfacial phonon modes (Figure 3c). The spectrum offers physical insights into the partial phonon density of states at/around the YBCO/LSAT interface, revealing clear differences with the behavior profiles obtained from YBCO and LSAT. The spatial sensitivity depends on the degree of scattering localization of the signal. It is worth highlighting that those modes are not accessible with traditional spectroscopy techniques that offer limited spatial sensitivity, which speaks to the scientific relevance of the current experimental findings of this work.

This analysis highlights the active role of surface contributions from the surrounding environment, thus requiring its inclusion for a more accurate interpretation of the complex EELS data. Our subtraction approach relies on the scattering physics and material properties, and the outcome still is dependent on the quality of the spectral subtraction and information obtained from other samples with similar characteristics, such as geometry and size. Values of the resultant spectrum should be considered as estimates that contain mostly interface scattering components with minor surface contributions from the surrounding and this is only valid at and near the interface.

To bring physical insights into the excitation of gamma-point longitudinal optical (LO) interfacial phonon modes, we calculated energy loss function (ELF) which accounts for the excitation of long wavelength bulk modes ($q \rightarrow 0$). Figure S3 shows the calculated ELF which displays three main peaks that lie within the experimental broad scattering peaks. These ELF peaks are associated with three interface bulk phonon modes involving lattice motion subjected to restoring forces induced from electrostatic fields from both YBCO and LSAT materials. Notice that the scattering associated with these types of modes is delocalized extending from the interface into both materials. Furthermore, we think that additional scattering associated with the excitations of short wavelength phonon modes also appears in the resultant spectrum (green curve in Fig. 3c), and they account for the differences between the experimental results in the calculated ELF. However, they cannot be distinguished from remaining surface contributions. Further calculations of scattering cross sections from lattice modes at interfaces accounting for impact scattering components ($q >> 0$) should shed light on the excitation of those phonon modes.

**D. Highly localized phonon response near the interface**

We probed phonons within the first few adjacent unit cells (below 4 nm) to investigate the local phonon behavior near and at the interface. Figure 4(a) depicts an atomically resolved HAADF-STEM image of the crystal structure at the YBCO/LSAT interface. The YBCO region appears in the upper half of the image while the LSAT substrate is at the bottom. The bright and dark bands correspond to the $CuO_2$ plane and CuO chain layers described in section A. The first pair of layers above the interface follow a Y-123 structure (where each digit is the number of Y, Ba, and Cu atoms, respectively), which transitions into a Y-124 structure usually characterized by the presence of double CuO chains (indicated by arrows in Fig. 4(a)). The interface forms between a brighter band at the top of the LSAT substrate and a Ba-terminated YBCO layer [3,16,17].

Figure 4(b) shows spatially resolved EELS maps acquired near the YBCO/LSAT interface for several energy loss regimes from 35 to 90 meV. We focused on the most prominent spectral features appearing on the YBCO side considering the scattering contributions found at the interface in our previous analysis (Fig. 3). Figure 4(c) shows the corresponding EEL spectra extracted at certain locations highlighting the vibrational response of unique structural features such as $CuO_2$ planes and CuO chains in YBCO.

The map for the 35-45 meV range shows a clear scattering modulation on the YBCO side but displays a smoother scattering profile on the LSAT side, which is in synchrony with the contrast pattern observed in the HAADF image (Fig. 4(a)). This oscillatory-type modulation exhibits enhancements in the region within the Y-123 layered units containing $CuO_2$ planes and reductions over the CuO chains, resulting in sub-nm regions of highly localized scattering within the YBCO. This suggests that phonon modes associated with the Y-123 layers (e.g. bending modes of $CuO_2$ planes) are responsible for the local 35-45 meV signal [19]. Note that a scattering peak at around 40 meV is also observed in our analysis of interface modes presented in Fig. 3(c), and the maps displays its spatial distribution near and at the interface. Furthermore, a close inspection of the corresponding EELS spectra (Fig. 4(c)) shows that the scattering modulation can be observed as an intensity variation of a shoulder-like feature which peaks at the $CuO_2$ plane positions.

This oscillatory scattering pattern distinguishing the layered Y-123 regions from CuO chains is still preserved in the neighboring 50-63 meV range, but with weaker intensity. Interestingly, the second layered $CuO_2$ plane displays a broader region of scattering localization likely due to contributions from excitations of the apical O modes (~ 55 meV) [19]. In a similar manner as found previously, this map also exhibits a weak signal at the interface, accompanied by a symmetric scattering profile of about 1 nm around it. The absence of scattering at the interface for the 35-45 meV and 50-63 meV bands suggests that a particular mode sustained in the closest Y-123 unit to the interface is evidently disrupted by the presence of the neighboring LSAT leading to the formation of a mode with distinct frequency.

The map for the 70-80 meV range reveals a highly localized scattering enhancement around the interface (light purple band), which contrasts with the absence of scattering observed in the lower energy maps. Note that these 70-80 meV contributions were also reported in our analysis presented in Fig. 3(c), and the map reveals its spatial distribution around the interface. Interestingly, we found that there is a drastic suppression of scattering in the region around the first Cu-O chain. However, the scattering observed in the second chain and first double chains are stronger and similar in intensity to the layered Y-123 regions generating a more uniform distribution above the first chain. This suppression behavior in the first Cu-O chain is unique and might be associated with the asymmetric atomic arrangement surrounding the chain imposed by the interface. Further experimental work is needed to map out if these atom-level scattering variations are preserved along the interface.

We also examine the neighboring upper energy range. The 80-90 meV map also reveals signal at the interface and maintain the whole scattering pattern as the 70-80 meV map. An analysis of the EELS spectra (Fig. 4(c)) reveals that peaks in this spectral range tend to broaden in the Cu-O chains positions, while at the layered Cu-O planes position the resonances get sharper. Those variations were not captured in the map, however, minor enhancement at regions near the $CuO_2$ planes 2 and 3 are noticed. Given that a LO-type interface bulk phonon mode is expected in this spectral range due to the YBCO and LSAT interaction (see Fig. S2), the roughly uniform intensity around the interface seems to suggest the delocalized excitation of this mode. Furthermore, anomalous modes (e.g. 85 meV Cu-O stretching mode) and hardening effects on optical modes due to local variations in oxygen-deficiency [24] might appear within this spectral range as spatially localized and delocalized contributions respectively, thus adding more complexity to the scattering maps.

Further analysis of the EELS spectra of Fig. 4(c) reveals differences between the single and the double Cu-O chains, the latter having a closer resemblance to a $CuO_2$ plane spectrum. We think that this might be attributed to a phonon hardening in the Y-124 structure, which brings the vibrational frequencies of the double Cu-O chains closer to those of the $CuO_2$ planes [24]. The signal on the LSAT region reveals a large contribution to the scattering in the investigated energy ranges but appears to be more delocalized. This is especially evident in the 90-100 meV range map (not shown) where signal only appears on the LSAT side. Furthermore, the LSAT spectra likewise reveal minimum variation within the first 5 nm away from the interface confirming the excitation modes at the interface.

We also compare the atomic-scale scattering behavior of the 40 and 75 meV modes near the interface with the measured scattering distribution of YBCO far from the interface. We found two main differences demonstrating that the scattering variations seen at close proximity to the interface are unique. The first difference is a shift of the position of the closest 40 meV scattering band to the interface. Figure S5(a) shows the band profile with respect to the ADF signal from the Y-123 unit structure. Notice that the position of the closest band is displaced towards the interface, while the other bands remained unaltered displaying the same pattern found in the Y-123 structures within the YBCO film. The second difference is revealed by a reduction of scattering at the 75 meV region observed at about 1.5 nm from the interface. These variations were not seen in YBCO regions distant from the interface, as illustrated in Figure S5(b). We think that those scattering variations are likely associated with local changes in the lattice vibrations due to its proximity to the LSAT support. These findings offer scientific motivation for theoretical work on EELS calculations at YBCO/LSAT interfaces which should extend our understanding of highly confined interfacial phonon modes in high Tc thin film cuprates.

In demonstrating the ability to resolve the spatial variation of phonon scattering near YBCO/LSAT interfaces, our atomic-scale EELS study provides a unique local probe of the electron-lattice interactions in cuprate superconductors, with the potential to shed new light on the high-Tc pairing mechanism. Since the cuprates exhibit doping-dependent inhomogeneities in both the charge and lattice degrees of freedom [25–27], as well as doping-dependent oxygen isotope effects on the various ordering temperatures [1], electron-phonon coupling is believed to play a crucial role in driving the various types of order, particularly the superconductivity. Specifically, high-Tc theories based on polaron formation and bipolaron pairing have offered salient explanations of experimental phenomena that challenge theories based solely on electron-mediated or spin-mediated pairing mechanism [28]. Our findings can crucially inform this ongoing theoretical debate, by revealing how the electron-phonon coupling in cuprates varies from the bulk to interfaces, in direct correlation with the local doping level and lattice symmetry, both of which being able to affect the various ordering phenomena at play.

# CONCLUSIONS

We unveiled a variety of phonon modes located at a YBCO/LSAT interface whose spectral response differ from those in bulk YBCO and LSAT. Some of those modes are linked unequivocally to Y-123 layered units (i.e. CuO planes), while the other modes can be linked to mixed contributions from YBCO and LSAT (i.e. interfacial phonon mode). We also demonstrated the significance of the surrounding environment to the scattering modulation of the acquired interface spectra due to excitation of surface phonon polaritons. We seek to inject caution into the acceptance of disregarding these contributions in the study of interface phonons and also provide a method to suppress and/or remove those long-range surface scattering contributions. This work provides key considerations for suitable interpretation of EELS scattering from layered heterostructures and uncovered lattice vibrational modes that are not usually considered in studies of high-$T_c$ superconducting cuprate films.

# ACKNOWLEDGMENTS

This work was supported by the Natural Sciences and Engineering Research Council of Canada (NSERC) under the Discovery Grant. M.J.L. acknowledges the support for infrastructure from the Canadian Foundation for Innovation (CFI) under the John R. Evans Leaders Fund (JELF) program and the Ontario Research Fund: Research Infrastructure (ORF-RI) program. M.J.L's research is undertaken in part thanks to funding from the Canada Research Chairs program. We also thank the Canadian Centre for Electron Microscopy (CCEM) for providing access to electron microscopy facilities. J.W thanks G.M. Zhao at CalState Los Angeles for helpful discussions.

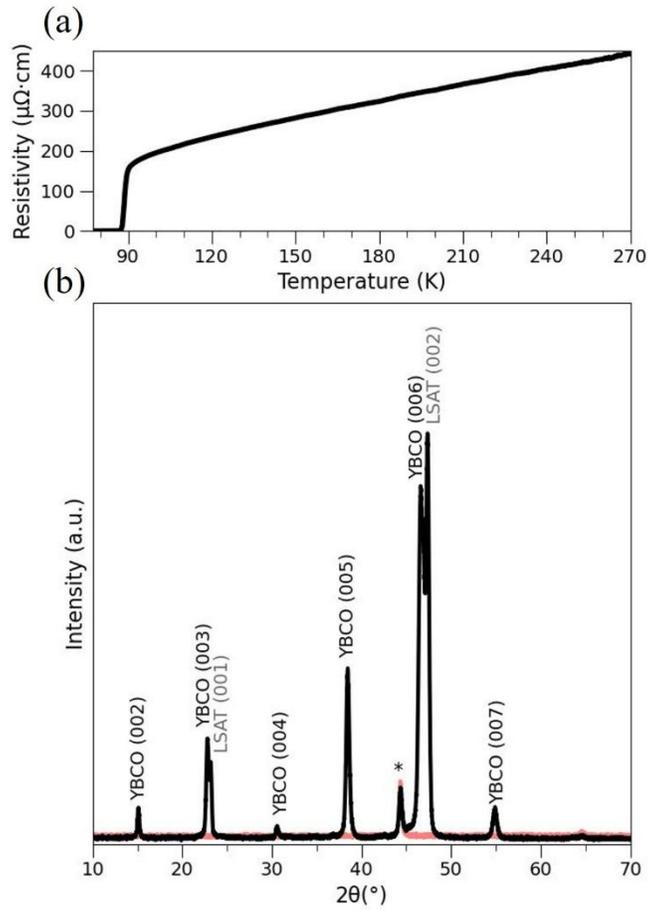

Figure 1. Superconductivity and structural properties of epitaxially-grown YBCO film. (a) Resistivity versus temperature measurement of the *c*-axis YBCO film showing a $T_c$ onset of about 90 K and transition width of about 3 K. (b) X-ray diffraction scans done in the θ-2θ mode of the c-axis YBCO film (black) and the background from the sample holder without the film (red). The peak labelled with (*) is attributed to the sample holder. All other prominent features are indexed as (00ℓ) peaks from the YBCO film or the LSAT substrate.

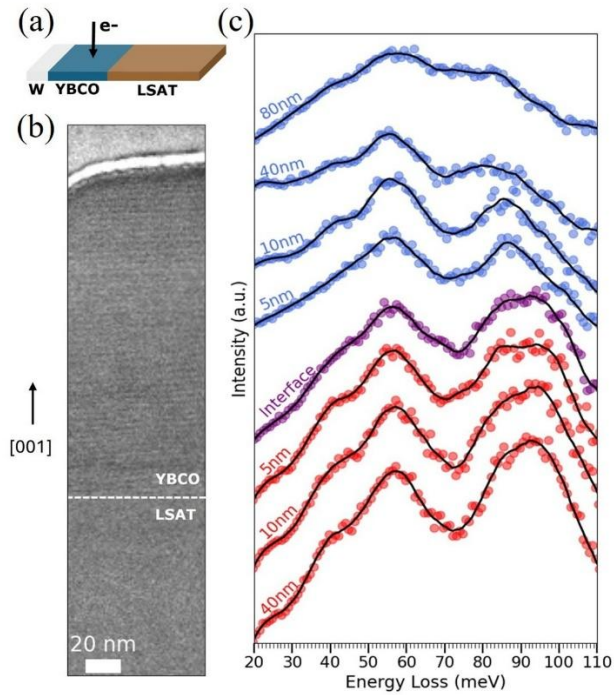

Figure 2. Phonon EELS behavior across the YBCO/LSAT interface. (a) Schematics of the fabricated cross-sectional YBCO/LSAT structure for the vibrational spectroscopy analysis. (a) HAADF-STEM image of the YBCO/LSAT interface. The interface is orthogonal to the [001] direction. (c) EELS spectra acquired at different positions across the interface. The spectra acquired at the interface appears in purple, while the ones acquired on the YBCO and LSAT structures are highlighted in red and blue respectively. Distances from the interface are indicated.

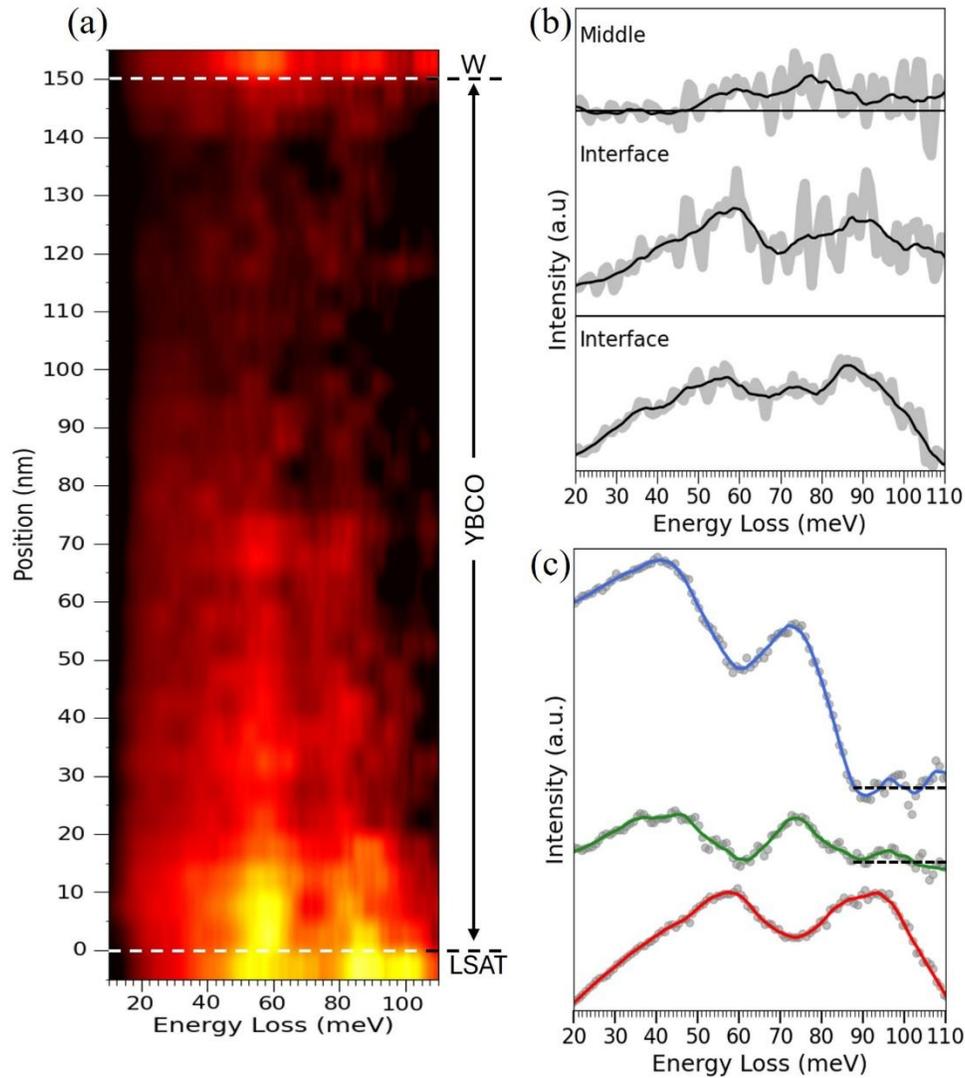

Figure 3. The role of surface phonon modes in interfacial phonon scattering. (a) A 2D projection of the EELS spectra acquired across the YBCO with 5 nm step. The color intensity represents the scattering intensity as a function of the energy loss (ΔE) at different positions. The boundaries of the YBCO are indicated with a horizontal dashed line. The spectral signal acquired in the LSAT and W layers are also included (b) Resultant subtracted spectra after removing the YBCO surface contributions. The resultant spectrum for the interface and middle positions are displayed at the top and middle. The noise gets amplified towards high energies due to effects of the multiplicative factor ($\Delta E^2$). An additional subtracted spectrum with lower noise level is also included for comparative purposes at the bottom. (c) Resultant subtracted EELS spectrum at the interface after removing both YBCO and LSAT contributions (green curve). Bulk YBCO (blue) and LSAT (red) phonon EELS spectra are shown for comparative purposes.

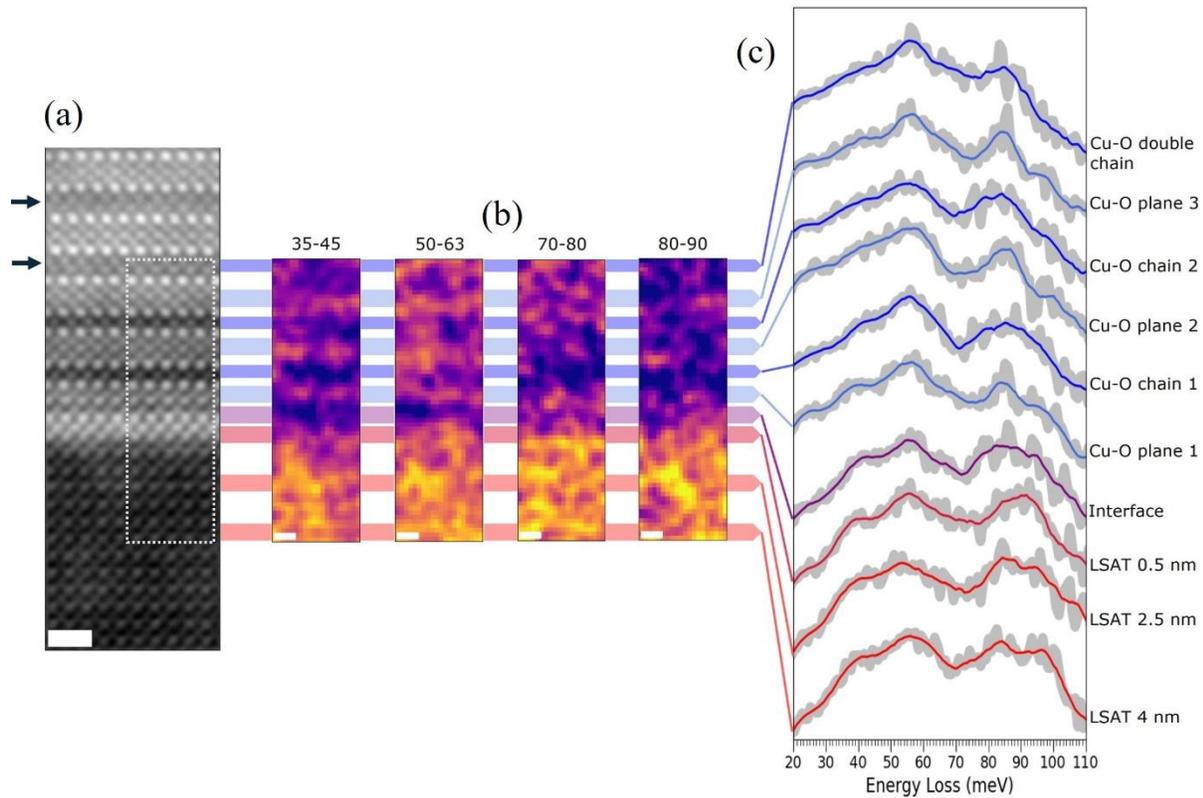

Figure 4. Localized phonon EELS response in the YBCO/LSAT interface. (a) Atomically resolved HAADF-STEM image of the YBCO/LSAT interface. Arrows indicate double CuO chains. Scale bar is 1 nm. (b) Spatially resolved EELS maps acquired from the rectangular area identified by a dotted line in the STEM image. Maps were integrated over spectral ranges linked to prominent spectral features in c). Scale bar is 0.5 nm. (c) EELS spectra acquired at different positions across the structure of the YBCO/LSAT interface. The position of the several CuO chains and $CuO_2$ planes on the YBCO are indicated. Color bands in the background serve as an aid to the eye to correlate the STEM image with the maps and spectra.